\documentclass{aa}
\usepackage{graphicx}
\usepackage{times}
\usepackage{natbib}
\usepackage[below]{placeins}
\usepackage{epsfig}
\usepackage{afterpage}
\usepackage{textcomp}

\newcommand{\sco}{\object{\mbox{Sco~X-1}}}
\newcommand{\mkn}{\object{Mrk~421}}

\newcommand{\chandra}{{\em Chandra}}

\newcommand\xmm{{\em \mbox{XMM-Newton}}}

\newcommand{\NH}{$N_{\rm H}$}
\newcommand{\HI}{ \ion{H}{i}}
\newcommand{\cms}{$\rm{cm}^{-2}$}
\newcommand{\green}{light grey}
\newcommand{\red}{dark grey}

\begin{document}

\titlerunning{EXAFS in the \sco\  spectrum}
\title{Physical properties of amorphous solid interstellar material from X-ray absorption spectroscopy of \sco }

\author{C.P. de Vries \inst{1} \and E. Costantini \inst{1,2}
	}

\offprints{C.P. de Vries, \\
	   \email{C.P.de.Vries@sron.nl}}

\institute{SRON, Netherlands Institute for Space Research, 
		Sorbonnelaan 2, 3584 CA Utrecht, The Netherlands
	   \and
	        Astronomical Institute, Utrecht University, 
	        PO Box 80000, 3508TA Utrecht, The Netherlands
	   }

\date{Received $\prec date1 \succ$, accepted  $\prec date2 \succ$}

\abstract{High quality, high resolution X-ray spectra were obtained of \sco\  with the
Reflection Grating Spectrometer (RGS) on board the \xmm\ satellite. 
The spectrum around the Oxygen K-edge is searched for signatures indicative 
for EXAFS. Analysis shows a clear indication for the existence of EXAFS with photo-electron
scattering distances in the absorbing medium in a range which applies to many solids. When scattering of the
photo-electron on Oxygen atoms is assumed, the Oxygen to Oxygen atom distances found are 2.75~\AA.
This fits with distances as found in amorphous water-ice, although water ice is thought to be an unlikely
constituent of the diffuse interstellar clouds which form the absorbing medium towards
\sco.   
}

\maketitle

\section{Introduction}
Extended X-ray absorption fine structures (EXAFS) refer to the oscillatory appearance
of X-ray spectra at the short wavelength side of an atomic edge. When an electron can
be emitted from an atom due to absorption of an X-ray photon, back-scattering
of the photo-electron wave by the surrounding atoms causes interferences
which may prevent the absorption of the X-ray photon in the first place. This process
is determined by the distances of the emitting atom to its neighbours. This interference 
behaviour will cause a sinusoidal structure of the absorption cross-section with wavelength.
Since the photo-electron will normally be absorbed within short distance by the medium into which
it is emitted, the EXAFS spectral structure will only show the average local surroundings of
emitting atoms. Whereas crystalline structures, which are repetitive over large distances, can be 
studied by X-ray diffraction techniques, EXAFS were recognized as an important tool to study
amorphous materials (see e.g. \cite{Sayers}). The availability of high intensity X-ray
facilities therefore has made the study of EXAFS an established technique in many fields of 
the materials sciences.

For astronomy, the availability of a new generation of large X-ray telescopes like \chandra\ and \xmm,
with their high resolution X-ray spectrometers, was already early recognized as an opportunity to observe
EXAFS of interstellar dust and hence study the character of the solid material of dust particles
(see. e.g. \cite{Woo}). However, until now only limited possible observations of interstellar EXAFS
have been reported (\cite{Petric}, \cite{Ueda}, \cite{jlee}).

The high intensity of the ground based X-ray sources used by the material sciences for EXAFS studies means that
signal to noise ratio is extremely high. In contrast, in astronomical sources noise
is a serious problem for EXAFS studies. In addition, since EXAFS extend over a broad range in wavelengths,
stability and knowledge of the instrument effective area and response is needed to a very high level
of accuracy.  



\begin{table*}[bhtp]
	\centering
	\caption[]{Overview of \xmm\ observations used in this article of \sco}
	\label{obs}
   \begin{tabular}{cccccl}
	Orbit & Observation & Date  & Duration (ksec) & pointing & CCD mode \\
	\hline 
	0224 & 0134550501 & 27-02-2001 &  8 x 2 & on-axis & single CCD \\ 
	0402 & 0153950201 & 18-02-2002 &  4 x 6 & off-axis & 2-CCD \\
	0592 & 0152890101 & 03-03-2003 &  15	& on-axis & RGS1 CCD4 \\
	     &		  &	       &  8 x 1.4 & on-axis & RGS2 single CCD \\
	     & 0152890201 & 	       &  9	& on-axis & RGS1 CCD4 \\
	     &		  &	       &  6 x 1.4 & on-axis & RGS2 single CCD \\
	     & 0152890301 &            &  9	& on-axis & RGS1 CCD4 \\
	     &            &            &  6 x 1.4 & on-axis & RGS2 single CCD \\
   \end{tabular}
\end{table*}

\begin{figure}[htbp]
 \resizebox{\hsize}{!}{\includegraphics[angle=90.0,clip]{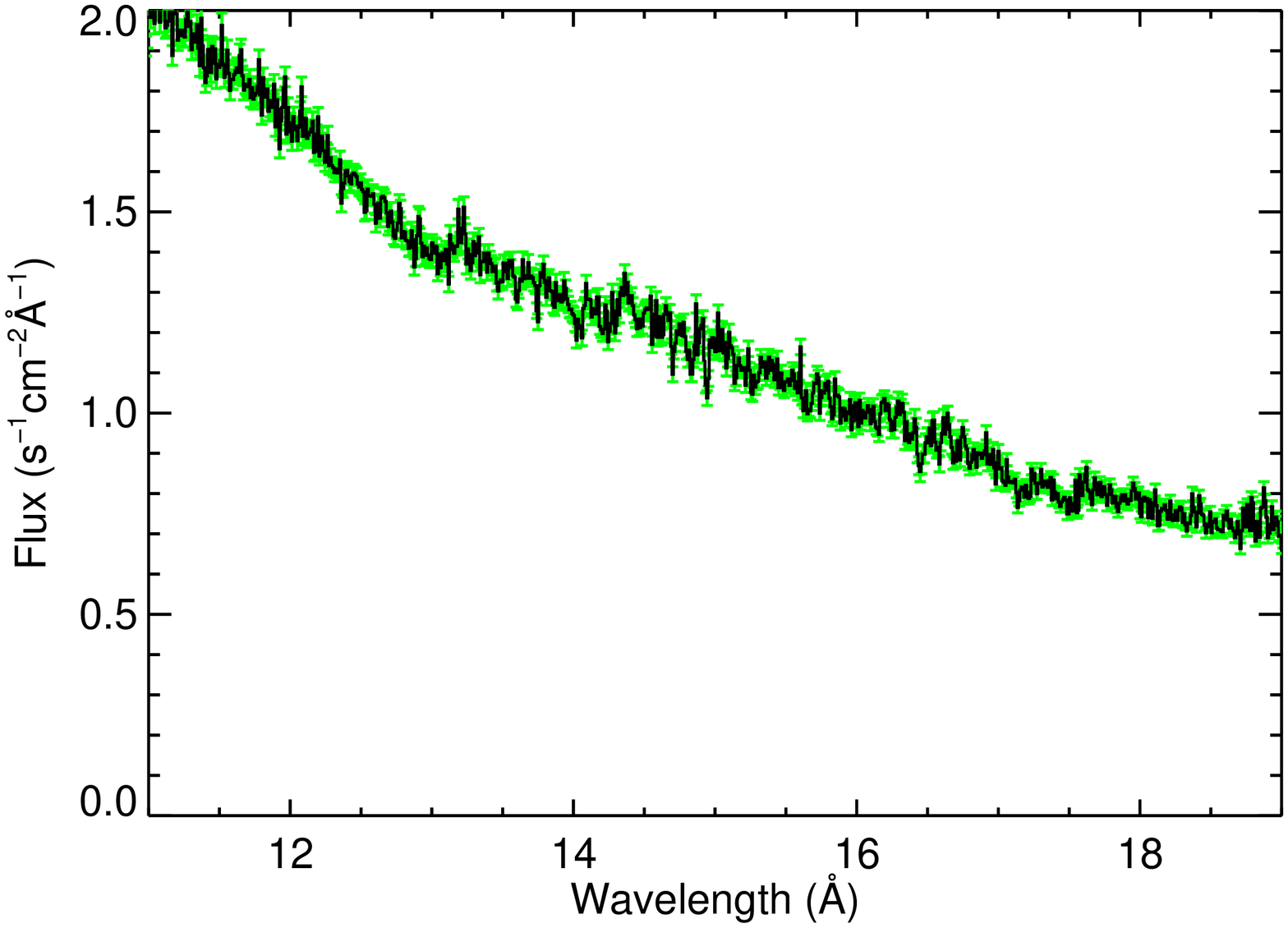}}
 \resizebox{\hsize}{!}{\includegraphics[angle=90.0,clip]{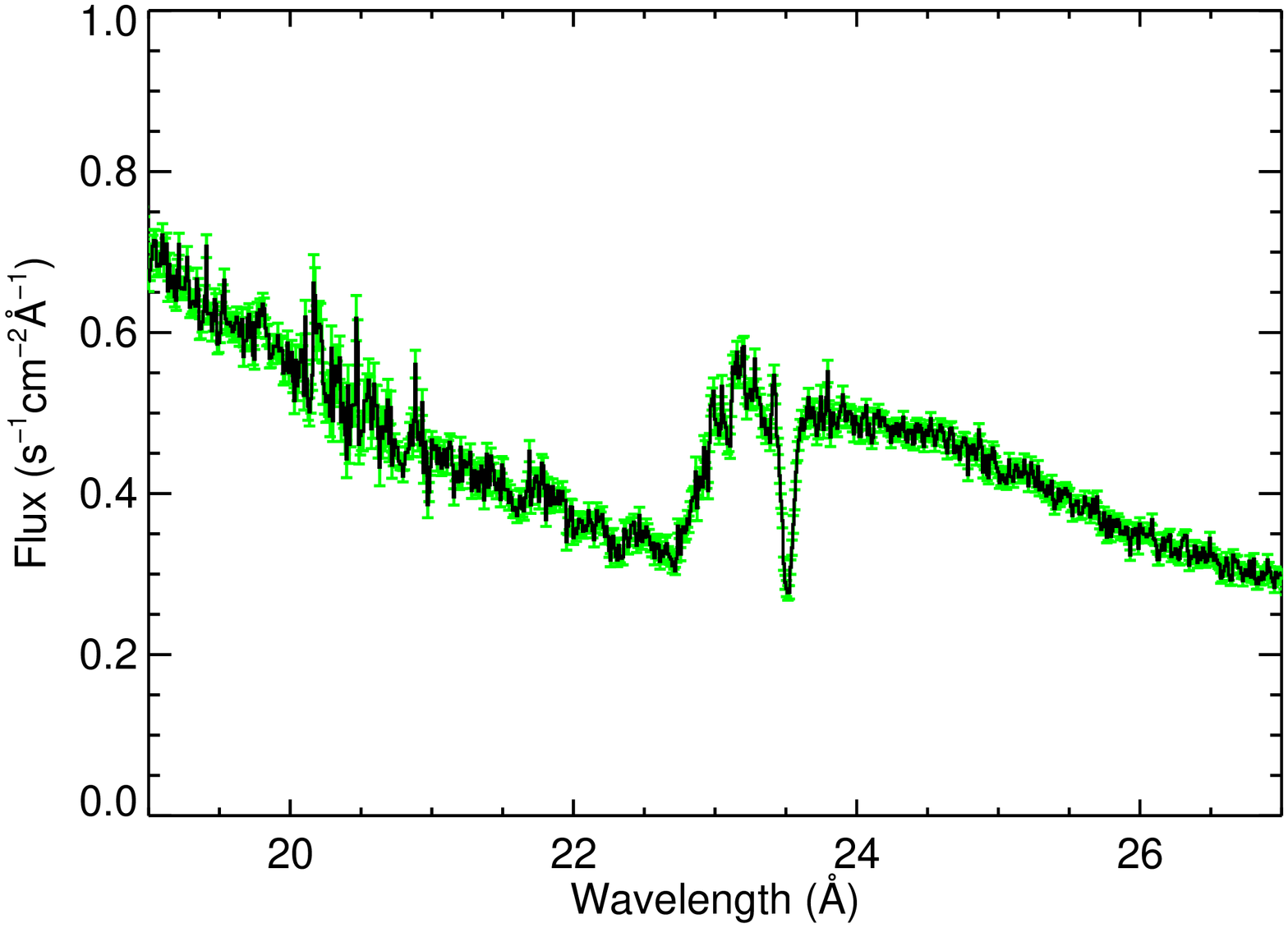}}
 \resizebox{\hsize}{!}{\includegraphics[angle=90.0,clip]{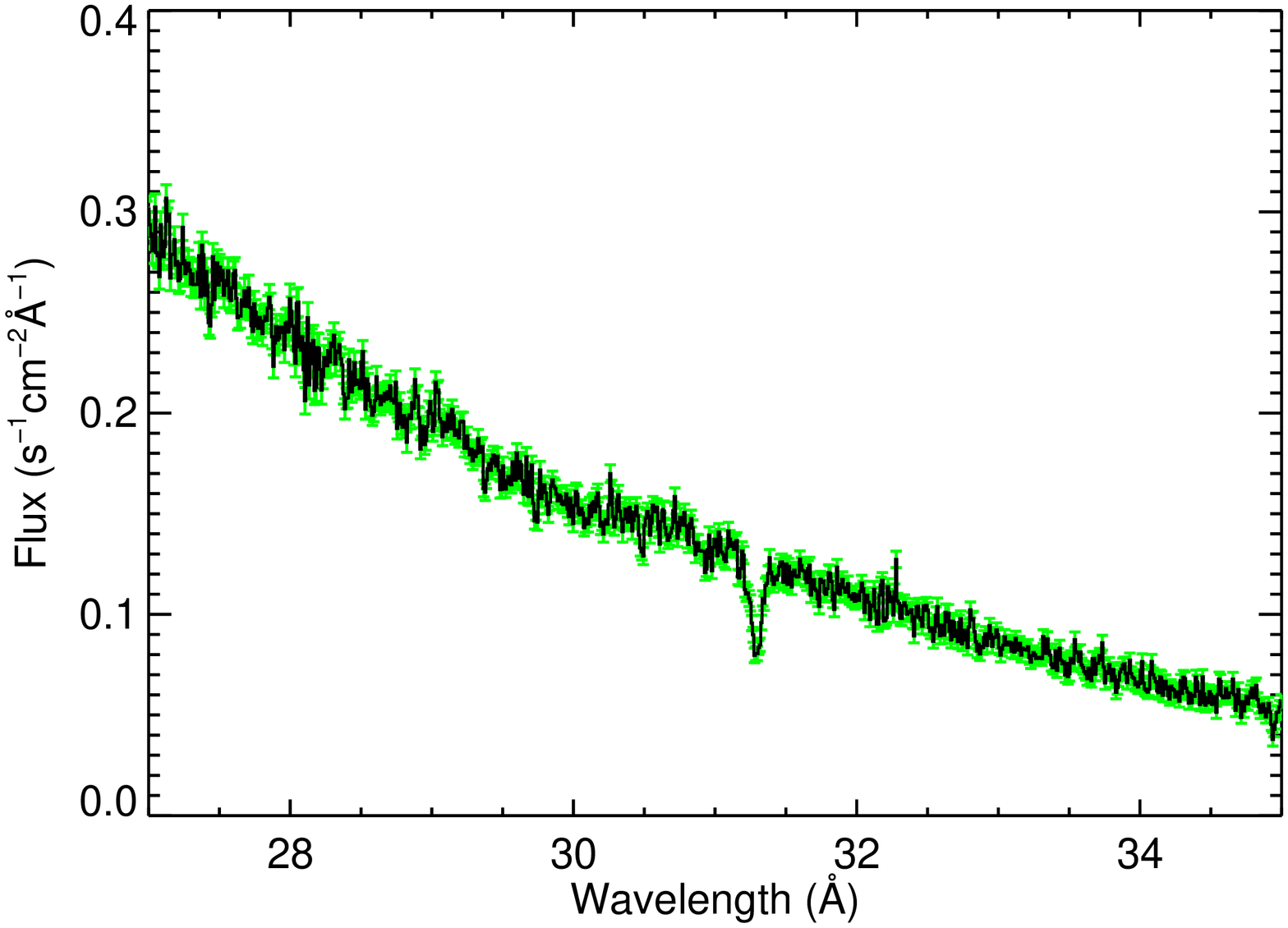}}	
 \caption{
	  Detailed view of different portions of the spectrum of \sco. The oxygen edge is easily
	  recognized at about 23~\AA, together with the 1s-2p absorption line of atomic oxygen
	  at 23.5~\AA. The nitrogen 1s-2p line is visible at 31.25~\AA. The somewhat increased
	  noise in the 20-21~\AA~ area is due to the fact that the wavelengths larger than 21~\AA~
	  had longer exposure times due to extra CCD4 exposures, while the offset observations
	  shifted the inoperable CCD to the 20-21~\AA~ area. The statistical error is plotted
	  as \green ~lines and amounts to about 2.5\% on average around the oxygen edge. 
 	 }
 \label{spect}
\end{figure}

\sco\ is one of the brightest X-ray sources in the sky and its emission is subject to interstellar 
absorption. It is located at a distance of about 2.8~kpc \citep{Bradshaw99} at a galactic latitude
of $23.8^\circ$, which means the source is situated about 1.1 kpc above the galactic plane. This is
well above the scale height of neutral interstellar material in the galaxy. For this reason the expected
total column density of \NH ~in front of the source can be derived from \HI~ surveys and is estimated
at about $19 \pm 3 \times 10^{20}$ \cms. This corresponds quite well with the \NH~ absorption derived
from low energy X-ray spectra \citep{deVries03}. Using a dust to gas ratio of 
$N_H/A_v=2.0 \times 10^{21}$~${\rm cm}^{-2}$ \citep{Vuong} we get a total of about 1 magnitude of extinction
at optical wavelengths. Comparing this with a general extinction of about 1.9~magnitude/kpc \citep{Savage} 
in the plane of the galaxy and keeping in mind that the line of sight to this
high latitude source only traverses
part of its total length through the molecular disk of the galaxy it is safe to assume that all low energy X-ray
absorption is due to the common diffuse material present in the galactic interstellar medium. 
In cold parts of this medium absorption by solid dust particles will play a significant role.

The high intensity of the X-ray flux from \sco\  provides high signal to noise ratio spectra in 
relatively short exposure times. This allows for a detailed study of the X-ray absorption characteristics
of the interstellar medium. In addition the high likelihood of absorption by solid matter makes this source ideally
suited to look for EXAFS.

\section{Observations and data reduction}

The \xmm ~RGS instrument \citep{herder} provides high resolution X-ray spectra in the soft energy
band (6 - 38~\AA). In this energy band the predominant interstellar absorption features are caused by
the elements oxygen, nitrogen, neon and iron (see Fig. \ref{spect}). Neon and nitrogen are thought
to be present mainly in atomic gas phase, while iron will be mainly chemically bound in compounds
contained in interstellar dust. 
Oxygen is one of the most abundant elements and causes the most prominent features in the spectrum, 
notably the deep Oxygen edge at around 23~\AA.
Oxygen is expected to be partly in (atomic) gas phase and partly bound in various chemical compounds in
solid form. Here we concentrate on signatures (EXAFS) of solids around the oxygen edge.

\sco\  has been observed on several occasions with the RGS instrument, in part for 
calibration purposes. Table \ref{obs} gives an overview of the observations used in this paper.
Due to the high X-ray flux of \sco, a standard spectroscopy mode observation would lead to 
unacceptable pile-up on the detector and loss of large amounts of observing time due to
overload of the on-board processing capabilities and limited telemetry capacity. For this
reason a faster readout mode was selected by reading individual single RGS CCD's in separate exposures.
The high flux of \sco\  also allowed for an off-axis observation mode, effectively shifting
the spectrum to a different location on the CCD's. At large offset angles (we used 27 arc-minutes),
the effective area of the instrument is reduced, decreasing potential problems with pile-up
thus allowing for multiple CCD readouts. This will increase effective exposure time.
Although large offset angles do introduce some additional uncertainty in absolute
flux, the change in shape of the response curve is mild and well-modeled. In the limited wavelength
range of interest (21-24~\AA) the effect is negligible and it is almost inconceivable that the off-axis
positions will introduce (or erase) a spectral modulation indicative of EXAFS.

Combining observations with different offsets
will decrease systematic errors in the spectra due to hot and erratic pixels on the CCD. Combining
observations with large offsets will also decrease small unknown fluctuations of effective
area due to inhomogeneous CCD response, effectively smoothing errors in the effective area. This will increase
the possibility of detecting real source broad wavelength flux fluctuations. The large offset
has the extra advantage that the important oxygen edge is covered
by both RGS instruments. In default on axis pointing mode the oxygen edge is only visible by one
RGS (RGS1) since the corresponding CCD on the other RGS has failed early in the mission. The large
offset shifted the oxygen edge to a different CCD location where both instruments have operational
CCD's.

In all, these special calibration mode observations have made this set of observations ideally
suited for the search for EXAFS.

Due to the shifted spectrum in the offset pointings, the total usable spectral range of the combined
observations extends from 12 to 38~\AA.

All data were reduced with the \xmm ~data analysis system SAS version 7. Since the source
was slightly variable between different observations and the scale of the off-axis effective area is 
not accurately known, obtained fluxes from the different observations have been scaled to
fit the revolution 0224. The average of the scaled, fluxed spectra (obtained using task "rgsfluxer")
was taken for further analysis (Fig. \ref{spect})

\section{Analysis and discussion}
Defining the edge energy at $E_0$ we can compute the wave number $k$ of the
escaping photo-electron as:

	\[ k = \frac{ 2 \pi}{h} \sqrt{ 2m ( \frac{hc}{\lambda} - E_0 ) } 
	\]

With $m$ the electron mass, $\lambda$ the wavelength and $h$ and $c$ Planck's constant and
the velocity of light respectively.

\begin{figure}[htbp]
 \resizebox{\hsize}{!}{\includegraphics[angle=90.0,clip]{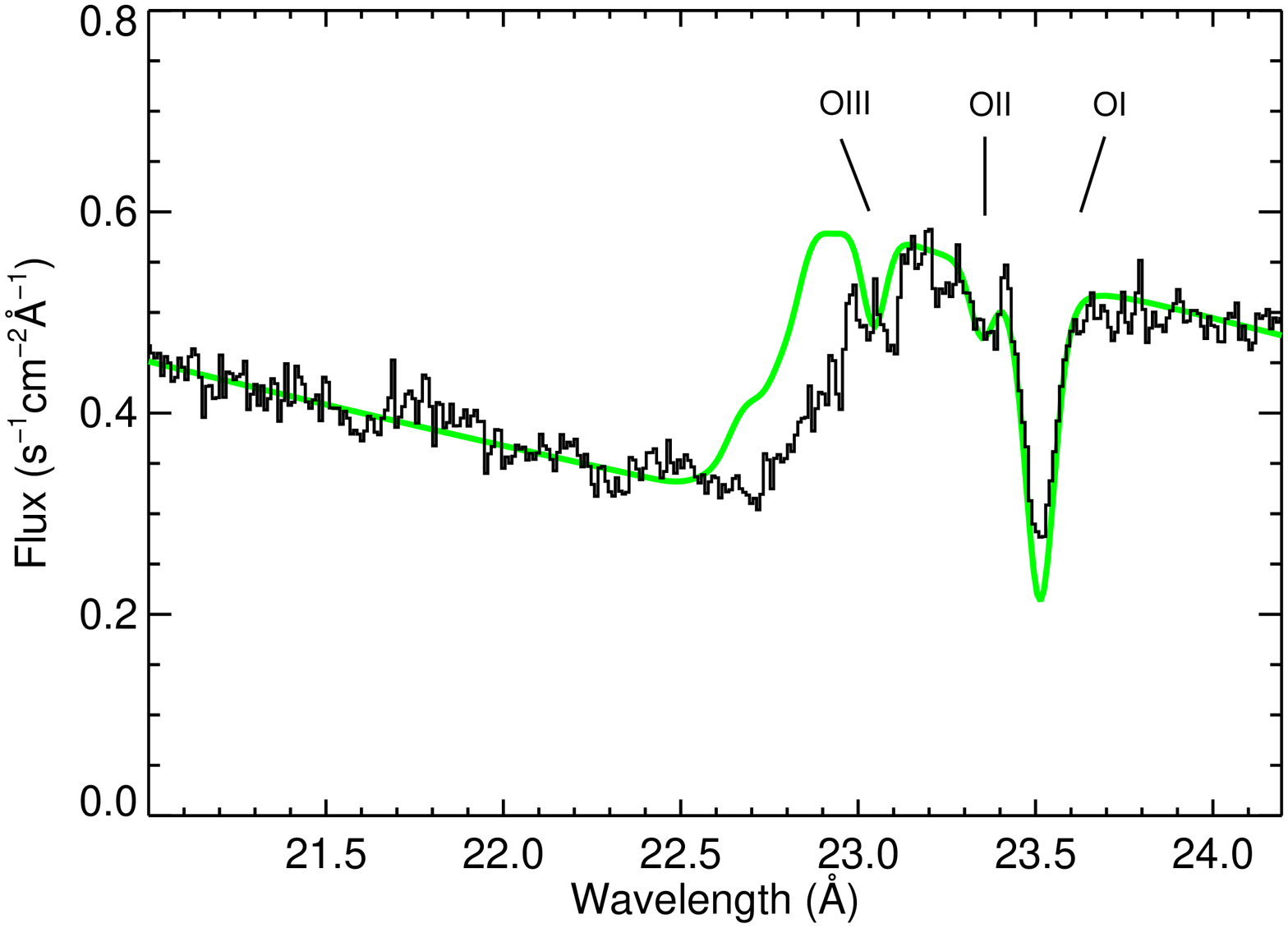}}
 \resizebox{\hsize}{!}{\includegraphics[angle=90.0,clip]{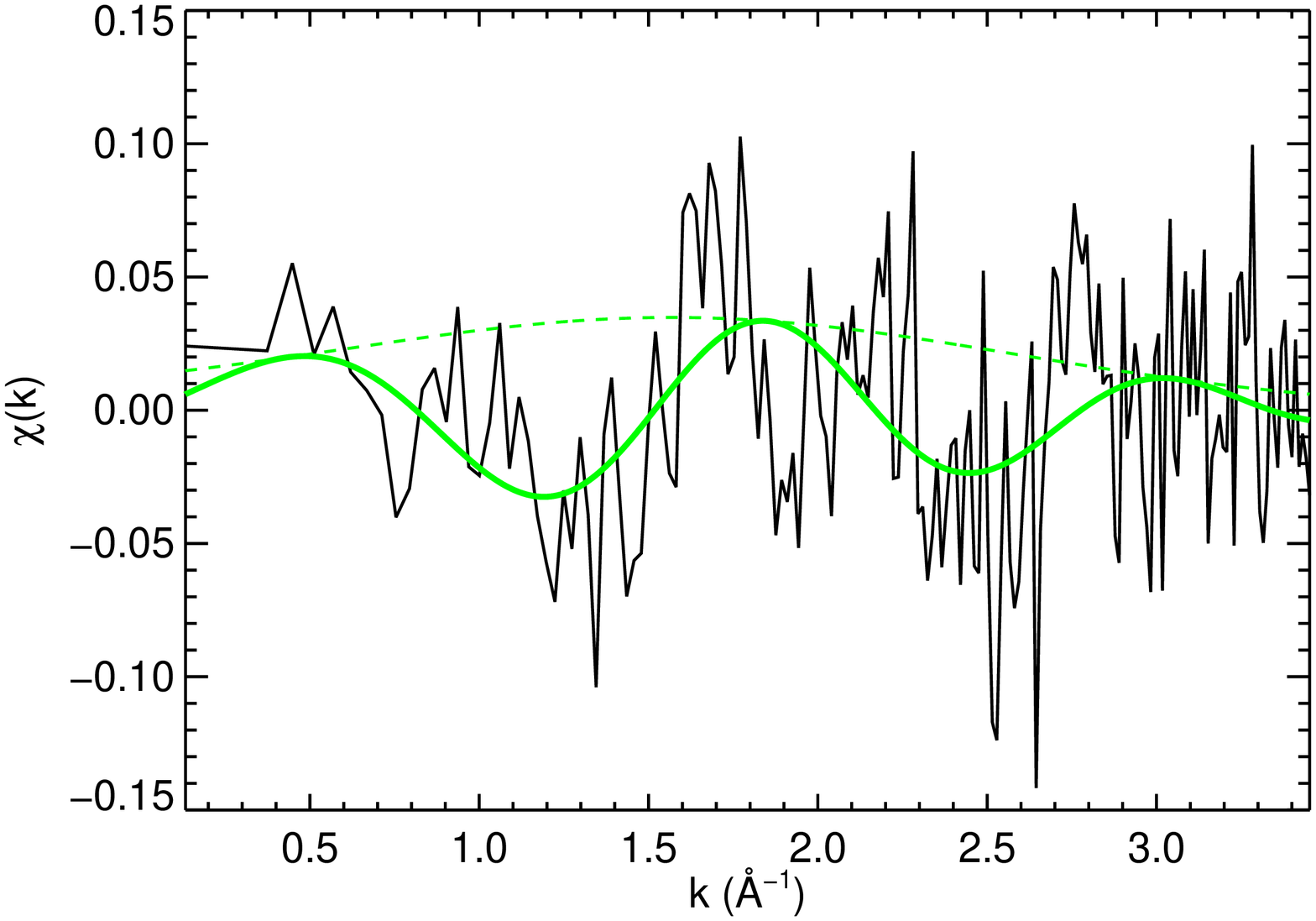}}
 \resizebox{\hsize}{!}{\includegraphics[angle=90.0,clip]{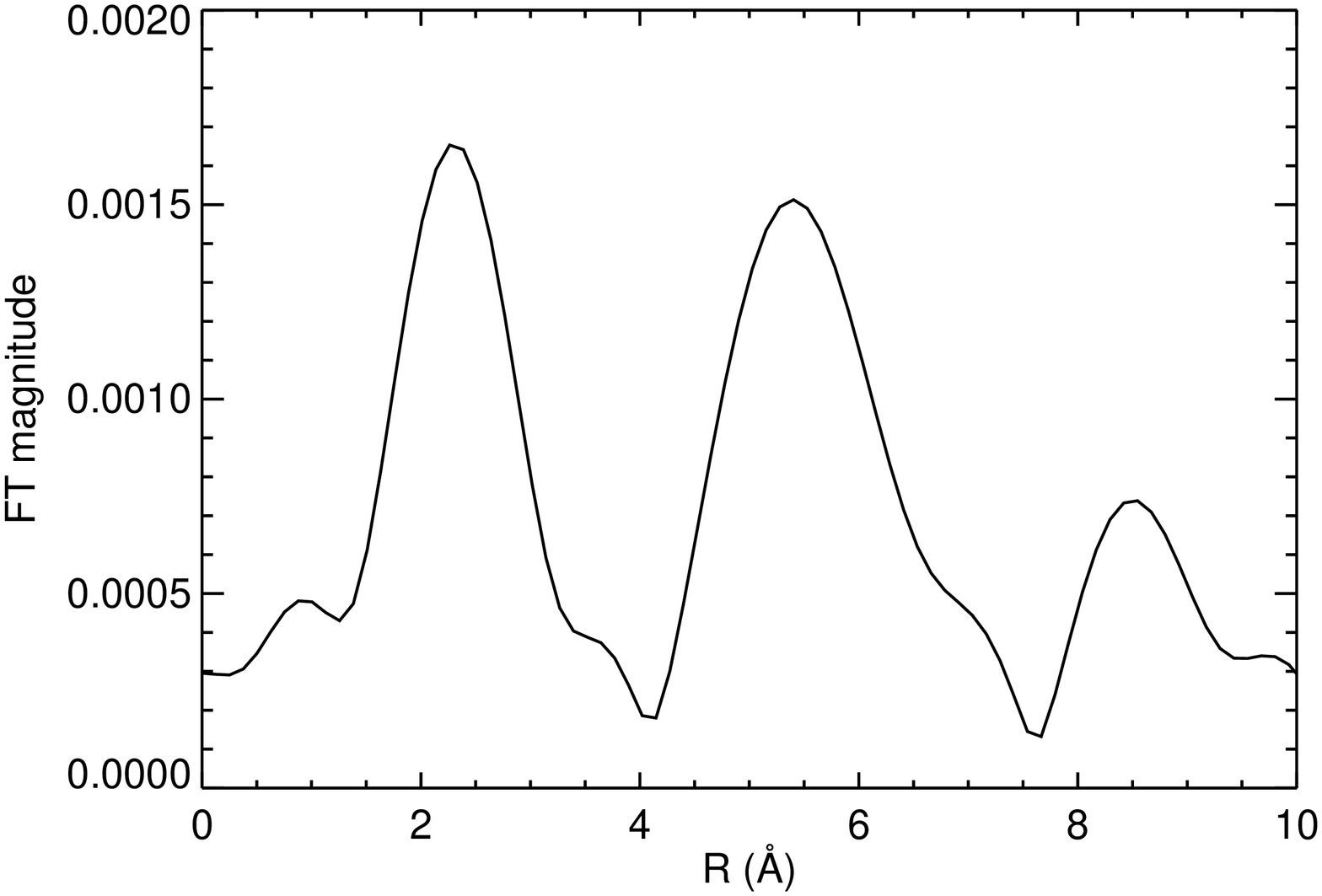}}	
 \caption{
	  EXAFS analysis of the \sco\  spectrum. Top plot shows the oxygen edge in the
	  spectrum. The \green ~line shows an indication for the expected smooth continuum based on the
	  atomic neutral oxygen absorption computed by \cite{mclaughlin}. The atomic 1s-2p resonance line
	  is indicated as well as possible additional resonance lines for
	  ${\rm O}_{\rm II}$ and ${\rm O}_{\rm III}$.  
	  Middle plot shows the absorption difference
	  with the continuum at energies above the edge (wavelengths
	  21.0-22.8~\AA) as function of $k$. The \green ~sinusoidal line shows the variations caused by the
	  main EXAFS peak at 2.4~\AA~in the FT magnitude plot (bottom). The dashed line traces
	  the amplitude of this fitted EXAFS feature. Bottom plot shows the absolute value of the
	  Fourier transform which traces the atomic scattering distances.
 	 }
 \label{esco}
\end{figure}

$\chi$ is defined as the relative change in the absorption coefficient $\mu$ with 
respect to the smooth continuum $\mu_0$ (see Fig. \ref{esco}, top) at energies above the edge:

	\[ \chi(k) = \frac{ ( \mu(k) - \mu_0(k) ) } { \mu_0(k) }
	\]

Because changes in the observed flux due to EXAFS are small with respect to the absolute flux, 
$\chi(k)$ can be obtained directly from the relative changes of the flux with respect to the
smooth continuum flux. This continuum is computed by taking the atomic oxygen absorption cross-section
as given by \cite{mclaughlin} applied to a linear slope across the small wavelength region of interest
(21-24~\AA ), fixing the continuum to the flux actually observed at the edges of the wavelength region. 
(see Fig. \ref{esco}, top plot)    

Plotting $\chi$ as function of $k$ (see Fig. \ref{esco}, middle plot), will reveal
EXAFS as a sinusoidal structure. Analyzing the spatial frequencies in this plot, by
taking the absolute value ($\sqrt{ {\rm (FT)} \cdot {\rm (FT)^*}}$) of the Fourier transform (FT), 
which scales linearly with EXAFS amplitudes, will show peaks representing the
distances of the scattering atoms in the absorbing solid. (The surface area under the peak corresponds
to the total amplitude of the EXAFS, since zeroes added to extend the $k$ range prior to
the transform (see e.g. \cite{Lee}) cause a smoothing (convolution) of the FT-magnitude graph). 
Peak positions ($R$) do not however
translate directly to atomic distances, due to phase shifts in the scattering process. 
Phase shifts must be known beforehand, e.g. from theoretical calculations, in order
to obtain true atomic distances.

A major problem is the adopted value for the edge energy $E_0$ \citep{Lee}. Although 
edge energies for many isolated atoms are known with sufficient precision, edge energies are
subject to chemical shifts for atoms bound in solids and are generally not known. Methods do
exist to resolve this issue by comparing the phase of the EXAFS (as function of $k$) with 
similar representative known solids. Since such data are lacking here, we adopted the strategy of
slightly varying $E_0$ within reasonable bounds and looking for features which remain constant.

In standard EXAFS analysis $\chi$ is multiplied with $k^3$ prior to Fourier transformation, to cancel
the decreasing power of EXAFS oscillations at large $k$. The extremely low noise in such (ground)
spectra does allow this weighting. In our spectrum however, noise is significant. Multiplying with
$k^3$ would blow up the Fourier transform at higher $k$ to unacceptable values. For this reason
$k^3$ multiplication was left out. The result is that the height and width of the peaks in the
power spectrum will not represent the true magnitude of the presence of a particular atomic distance
and scattering cross-section in the solid. For this reason we will only discuss the position of the
peak (and hence distances) found and not their magnitude.

Figure \ref{esco} shows the EXAFS analysis described above for \sco. The top plot shows the spectrum in the
wavelength region of interest, including the smooth continuum based on the atomic oxygen cross sections
computed by \cite{mclaughlin}. These cross sections include the 1s-2p \ion{O}{i} resonance feature
indicated in the plot. The fact that comparing the combined fit of edge depth and line shows a shallower line in the
actual spectrum may be due to the fact that not all oxygen is in atomic form. Bound oxygen may
have a strongly diminished 1s-2p feature when the 2p shell can be thought of as being filled by the chemical binding.
The edge depth on the other hand does sample both oxygen in bound and atomic state. In addition possible
resonance lines of \ion{O}{ii} and \ion{O}{iii} are fitted (see e.g. \cite{costantini})
These resonance structures are
indicated for reference only and have no meaning for the EXAFS analysis. The apparent shift of the observed
edge structure with respect to the computed atomic oxygen edge may also be due to the fact that part of the oxygen
is bound in solids. A detailed analysis of the edge structure itself however is beyond the scope of this
paper and will be discussed in a subsequent article.

\begin{figure}[htbp]
\resizebox{\hsize}{!}{\includegraphics[angle=90.0,clip]{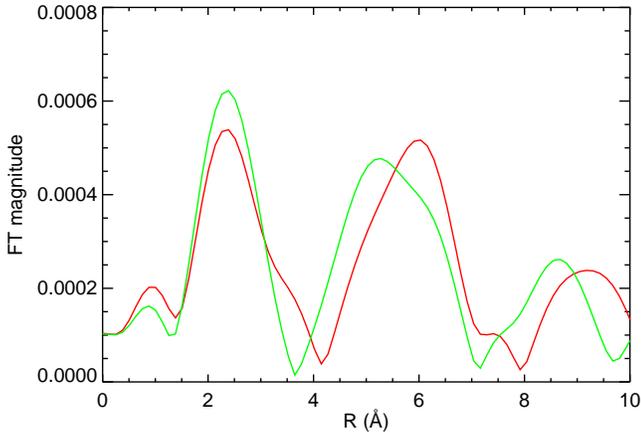}}
\caption{Variations in the FT magnitude when varying the edge energy $E_0$ between 22.70 (\green ~line) and
	 22.80~\AA (\red ~line). The 2.40~\AA~ peak remains on its position, while the other peaks change
	 somewhat.
	}
\label{ftvar}
\end{figure}

For our EXAFS analysis, the edge energy was varied between
22.6 and 22.9~\AA~ and a representative plot with $E_0=22.75$ is shown in figure \ref{esco} (middle plot). 
Looking at the peaks in 
the FT magnitude, it is found that the peak at $R=2.4$~\AA~ is quite insensitive to changes in edge energy
and varies only by about 0.1~\AA~ within an appropriate range of edge energies ($E_0=$22.70 - 22.80~\AA), 
while the peak at $R=5.5$~\AA~ is a bit more variable in both magnitude and position (see figure \ref{ftvar})
Selecting the $R=2.4$~\AA~ peak and Fourier transforming back to the $k$ scale, we get the sinusoidal curve
(\green~line) in the middle plot. This clearly shows that the selected spatial frequency indeed follows
the structure of the raw data and that sinusoidal signatures do exist in the data. This is clearly an
indication for the presence of EXAFS in the data. 

Next issue is to see if these EXAFS are originating in the source, or can be attributed to 
instrumental effects. For this reason the same analysis was performed on the total of all
RGS \mkn\  data. All these data-sets combine a total of about 1.3 Ms of exposure time. The line
of sight towards \mkn\  suffers
only from very low extinction \citep{deVries03} and no significant absorption by solid material
is expected there. As in the
case of \sco\  offset observations were performed, but in this case the magnitude of the offsets
was limited. For the oxygen edge therefore only one RGS can be used. The limited offsets also
mean that possible gradual unknown errors in effective area will diminish less than in the case
of the \sco\  spectrum.

\begin{figure}[htbp]
 \resizebox{\hsize}{!}{\includegraphics[angle=90.0,clip]{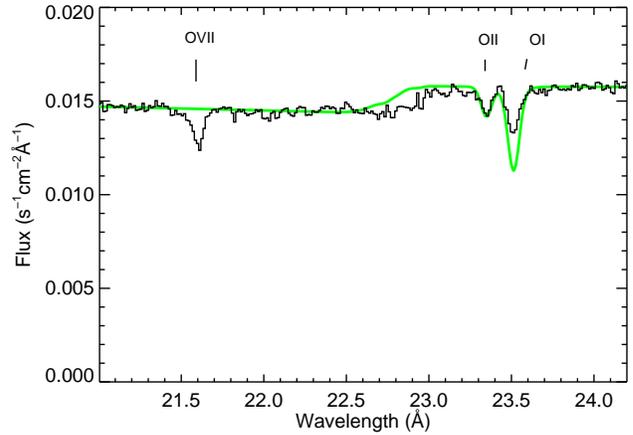}}
 \resizebox{\hsize}{!}{\includegraphics[angle=90.0,clip]{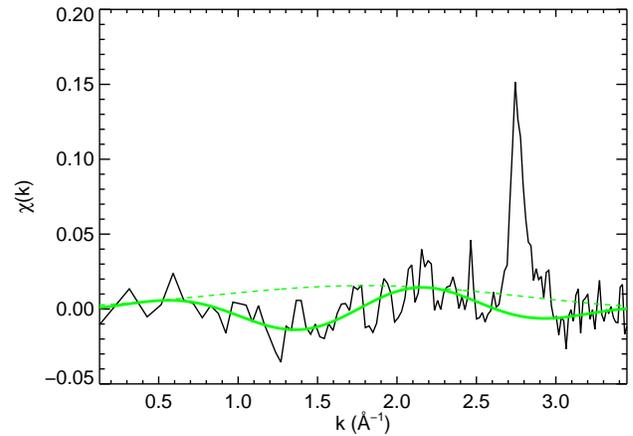}}
 \resizebox{\hsize}{!}{\includegraphics[angle=90.0,clip]{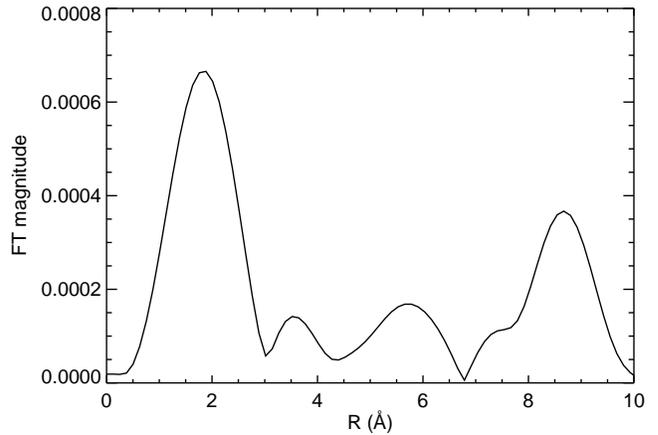}}	
 \caption{
	  EXAFS analysis of the \mkn\  spectrum for comparison. Plots are similar as in figure 
	  \ref{esco}. In addition to the ${\rm O}_{\rm I}$ and ${\rm O}_{\rm II}$ resonance
	  features the intergalactic ${\rm O}_{\rm VII}$ absorption is visible. This line 
	  is excluded from the analysis.
	  The first major peak in the FT magnitude plot (bottom) of 2.4~\AA\  of the \sco\ 
	  spectrum, in this \mkn\ plot is shifted to a lower R at 1.85~\AA~and has only
	  40\% of the \sco\  intensity, when the relative 
	  difference in source flux is taken into account.
 	 }
 \label{emkn}
\end{figure}

In figure \ref{emkn} the result is shown. A first peak does appear at 1.9~\AA, which is close to the
2.4~\AA~ position of the \sco\  spectrum. However, comparing the FT magnitudes of figures \ref{esco}
and \ref{emkn} (bottom plots) it can 
be seen that the FT magnitude of this peak as derived from $\chi(k)$
is only 40\% of the 2.4~\AA~ peak of \sco. $\chi(k)$ is defined as the relative change in absorption 
which corrects for the relative difference between source fluxes. Since peak widths are virtually 
identical this also means the surface of the \mkn\  peak, which traces the total EXAFS magnitude,
is 40\% of the \sco\  feature. This is confirmed by proper analysis of the peaks. 
This difference is significant, since the error in the amplitude, derived from the average
statistical error around the oxygen edge and the number of datapoints used in the fit, was
computed at about 10\% only for both peaks.

It appears that the source with largest absorption (\sco) does have the largest EXAFS signature. For purely
instrumental features, one would have expected equal signatures.

\begin{figure}[htbp]
\resizebox{\hsize}{!}{\includegraphics[angle=90.0,clip]{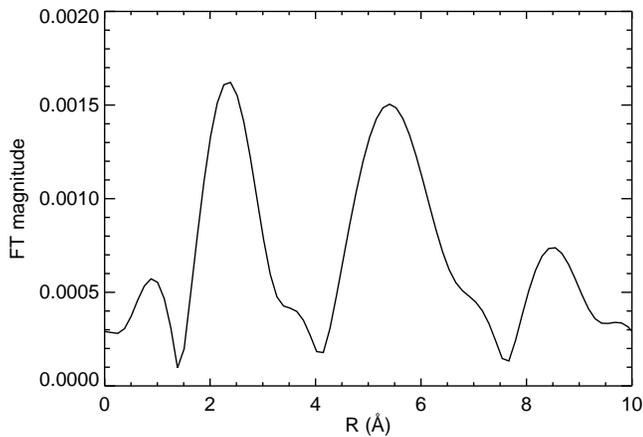}}
\caption{
	 Fourier transform of the \sco\  $\chi(k)$ data (Fig. \ref{esco}, middle plot) 
	 with the \mkn\  EXAFS fit, which contains possible instrumental features
	 (Fig. \ref{emkn}, middle plot), removed. 
	The basic features of Fig. \ref{esco} remain, confirming their likely true
	interstellar origin.
	}
 \label{sco-mkn}
\end{figure}

An extra check was made by removing the EXAFS found in \mkn, which may trace instrumental
features, from the derived $\chi(k)$ in \sco\ by subtracting the relative EXAFS amplitudes (relative to the 
smooth continuum and keeping
in mind that the amplitudes of the EXAFS are small with respect to the total flux) and
check the resulting Fourier transform. This plot is shown in Fig. \ref{sco-mkn}. 
The main peaks of Fig. \ref{esco} remain, confirming their likely interstellar nature.
The top amplitudes in the Fourier transform hardly change, since the peak in the \mkn\ 
Fourier transform was somewhat shifted with respect to the peak in the \sco\  FT transform and
phases of $\chi(k)$ slightly differ.
The shape of the R=2.40~\AA~ peak changes in the sense that it becomes somewhat more narrow
on the short R side. 

The peak positions found are in a range of distances (taken into account possible phase shifts
up to 0.4~\AA) which apply to a multitude of minerals. Unfortunately, proper laboratory data on the
solid structures expected for interstellar dust grains are scarce. The data presented here can
help solving the likely structure of solid dust grains when data on appropriate compounds
become available.  
 
A likely candidate atom for the photo-electron scattering is another oxygen atom, since those
are very abundant. Phase shifts for scattering on oxygen atoms are computed at about 0.35~\AA~  
\citep{Zubavichus}. Therefore the atomic distance probed by a peak at R=2.40~\AA~ amounts to a
real distance of 2.75~\AA. Amorphous water-ice is known to have an oxygen-to-oxygen distance of 2.73~\AA~  
\citep{Zubavichus}, which fits very nicely with the peak we find, given our peak-position
uncertainty of about 0.1~\AA. However, other compounds with electron scattering on other atoms then
oxygen cannot be ruled out. \cite{Woo} reports a set of atomic distances for different atoms
for different minerals. None of these distances seem to be observed in here, but the list is by no
means exhaustive. Furthermore, not much is known about the exact mineral atomic lattice structure in dust
grains which might be different from the solids generally found on earth.  

It is known that the RGS instrument features an instrumental oxygen edge, which is constant with time and
is taken into account by the calibration during data processing. This instrumental oxygen,
which amounts to an Oxygen column depth of about $N_O= 2 \times 10^{17} \: {\rm cm}^{-2}$ is
most likely in the form of a thin ice-layer of about 550~\AA~ on top of the detector \citep{deVries03}.
This ice will also cause EXAFS features which are not taken into account by the calibrations. The $R=1.9$~\AA~
feature in the \mkn\ spectrum may be due to a combination of this ice layer and small uncorrected wiggles
in the effective area. As such the intensity of the $R=1.9$~\AA~ peak can thus be regarded as an upper
limit for EXAFS caused by the instrumental ice layer. If we assume the \sco\  EXAFS peak to be due to
ice, the fact that the peak for \mkn\ is at a somewhat
different position (and phase) might be either due to the interference of residual instrumental
effects, or by the fact that the instrumental oxygen is bound in a different form of ice compared to the possible
ice on interstellar dust grains (e.g. a crystal lattice compared to amorphous ice). Of course the \sco\  EXAFS
peak may represent a different mineral than ice.

The \sco\  EXAFS will also contain instrumental features present. Since the thickness of the instrumental ice layer is constant with time, as checked by regular calibrations, and the \sco observations were performed in the same
time frame as the \mkn observations, the spectra of both sources will be similarly affected by the instrumental
ice layer. Assuming that the amplitude of both
\sco\  and \mkn\  EXAFS peaks scale in the same way with the amount of oxygen, this means that a maximum of 40\% of
the \sco\  EXAFS $R=2.4$~\AA~ peak may be instrumental. If we assume the total instrumental EXAFS feature
$R=1.9$~\AA~ peak to be due to an Oxygen column density of $N_O= 2 \times 10^{17} {\rm cm}^{-2}$, the
interstellar component of the \sco\  EXAFS, a minimum of 60\% of its total peak, would amount to 1.5 times the
instrumental Oxygen which is $N_O= 3 \times 10^{17} \: {\rm cm}^{-2}$. Since the total interstellar Oxygen
column density towards \sco\  is about $N_O= 10 \times 10^{17} \: {\rm cm}^{-2}$ \citep{deVries03}, this means
that the intensity of the \sco\  EXAFS spectral features hints that at least 30\% of the interstellar Oxygen
in the galactic plane towards \sco\  is bound in solids, if we assume that the peak intensity in the EXAFS
Fourier plot behaves as it would do if it would represent ice. If the total EXAFS signal in \sco\  
can be attributed to interstellar origin, this would present an upper limit of 50\% bound in solids.

This 30-50\% part for solids is only a very approximate number given instrumental uncertainties and 
other systematics and should only be seen as an indication.
However, such a number does not appear entirely unrealistic. A major part of the Oxygen will be in atomic
form (given the visibility of the narrow 1s-2p line) and not be part of solids. 
If other atoms than oxygen are responsible for the photo-electron back scatter, as can be the case
in other solids than ice, these atoms can have higher cross-sections for the back scatter process. In such
cases the estimate for the amount of oxygen bound in solids will be correspondingly lower. 

The EXAFS peak does fit remarkably well with amorphous ice.
Problem however is that the extinction towards \sco\ is likely caused by diffuse dust clouds in the
general interstellar medium. Although water-ice is present in the cores of dense molecular clouds, it is
hardly found in diffuse clouds (see e.g. \cite{Draine05}). Other, more robust minerals containing
Oxygen, like silicates, are thought to
form dust particles in diffuse clouds. Although water-ice, given the position found for our
peak in the EXAFS plot, does fit, we need data on other minerals to see if water-ice
remains the only candidate. Especially around the Oxygen edge however, such data are scarce.  

\section{Conclusions}
In the X-ray spectrum of \sco, the Oxygen-K edge was searched for the existence of EXAFS, which indicate
the presence of solids in the absorbing medium. A clear signal was found. Comparing with spectra of \mkn\ 
it is found that instrumental effects may account for 40\% of the signal. Assuming that EXAFS signals
scale with that of water-ice we find roughly 30-50\% of the absorbing oxygen is bound
in solid material. Although amorphous water-ice does fit the EXAFS peaks found, this material is an
unlikely candidate given the diffuse character of the absorbing medium. The data presented here can help
solving the character of interstellar dust grains when appropriate laboratory data on various plausible
materials become available.   

\begin{acknowledgements}
Based on observations obtained with XMM-Newton, an ESA science mission with instruments and contributions
directly funded by ESA Member States and NASA. The authors want to thank J. Kaastra for reading of the manuscript. 
\end{acknowledgements}

\bibliographystyle{aa}
\bibliography{ref}

\end{document}